# Control of proton transport and hydrogenation in double-gated graphene

J. Tong[1,2], Y. Fu[1,2], D. Domaretskiy[1], F. Della Pia[3], P. Dagar[1,2], L. Powell[1], D. Bahamon[4,5], S. Huang[1,2], B. Xin[1,2], R. N. Costa Filho[6], L. F. Vega[4,5], I. V. Grigorieva[1], F. M. Peeters[6,7], A. Michaelides[3], M. Lozada-Hidalgo[1,2]

[1] Department of Physics and Astronomy, The University of Manchester, Manchester M13 9PL, UK
[2] National Graphene Institute, The University of Manchester, Manchester M13 9PL, UK
[3] Yusuf Hamied Department of Chemistry, University of Cambridge, Cambridge, CB2 1EW, U.K.
[4] Research and Innovation Center on CO2 and Hydrogen (RICH Center) and Chemical Engineering Department, Khalifa University, PO Box 127788, Abu Dhabi, United Arab Emirates
[5] Research and Innovation Center for graphene and 2D materials (RIC2D), Khalifa University, PO Box 127788, Abu Dhabi, United Arab Emirates
[6] Departamento de Física, Universidade Federal do Ceará, 60455-900 Fortaleza, Ceará, Brazil
[7] Departement Fysica, Universiteit Antwerpen, Groenenborgerlaan 171, B-2020 Antwerp, Belgium
[+] These authors contributed equally to this work.

**Graphene's basal plane can function as a perfectly selective barrier permeable to protons[1,2] but impermeable to all ions[3,4] and gases[5,6], stimulating its use in applications such as membranes[1,2,7,8], catalysis[9,10] and isotope separation[11,12]. Protons can also chemically adsorb on graphene and hydrogenate it[13,14], inducing a conductor-insulator transition intensely explored in graphene electronic devices[13-17]. However, both processes face energy barriers[1,12,18] that in the case of proton transport motivate strategies to accelerate it, such as introducing vacancies[4,7,8], incorporating catalytic metals[1,19] or chemically functionalising the lattice[18,20], but these can compromise other properties like ion selectivity[21,22] or mechanical stability[23]. Here we show that independent control of the electric field $E \sim$ V nm$^{-1}$ and charge carrier density $n \sim 10^{14}$ cm$^{-2}$ in double gated graphene allows decoupling proton transport from lattice hydrogenation and can accelerate proton transport such that it approaches the limiting electrolyte current in our devices. Proton transport and hydrogenation can be driven selectively with precision and robustness that enables proton-based logic-and-memory graphene devices with orders-of-magnitude on-off ratios. Our results show that field effects can accelerate and decouple electrochemical processes in double-gated 2D crystals and demonstrate the possibility of mapping such processes as a function of $E$ and $n$ – a fundamentally different technique to study 2D electrode-electrolyte interfaces.**

The charge density, $n$, and the electric field perpendicular to an electrode-electrolyte interface, $E$, are fundamentally linked via the applied electrical potential and experimental conditions such as ion concentration or solvent polarisability[24]. In contrast, electron transport studies have established that electrostatically gating a 2D crystal on both of its surfaces, so-called double gating, using either crystalline dielectrics[25-28] or liquid electrolytes[29-31] enables decoupling $E$ and $n$ because the individual gate potentials superpose in the 2D crystal. The independent control of these variables in 2D electronic transport devices[25-30] is now being routinely used to modify the band structure of 2D crystals, for example quenching the band gap of 2D semiconductors[29,30], or to enable precise electrostatic control of phases such as coupled ferroelectricity and superconductivity[26]. In this context, we hypothesise that double-gating could enable the study of proton transport[1,2] and hydrogenation[13,14] in graphene with independent control of $E$ and $n$, which is not currently possible. Pristine graphene, impermeable to all atoms and molecules at ambient conditions[5,6], is permeable to thermal protons in the direction perpendicular to its basal plane[1,2]. While it was suggested that pinholes in the lattice were necessary for the transport, recent work[2] discarded this interpretation, revealing instead that strain and curvature within wrinkles and nano-ripples intrinsic to the crystal lower the energy barrier for the transport. On the other hand, the hydrogenation of graphene[13,15], originally studied using hydrogen plasmas[13,32,33], was recently shown to proceed efficiently in an electrochemical setup using a non-aqueous electrolyte[14]. This process is characterised by a reversible but hysteretic gate-controlled conductor-insulator transition in graphene[14] – accompanied by a prominent $D$-band in its Raman spectrum – which arises when electrons and protons combine to yield adsorbed H atoms in graphene. In this work we study these two well-known electrochemical processes with independent control of $E$ and $n$ in graphene and find that this enables these processes to be driven with otherwise unattainable selectivity.

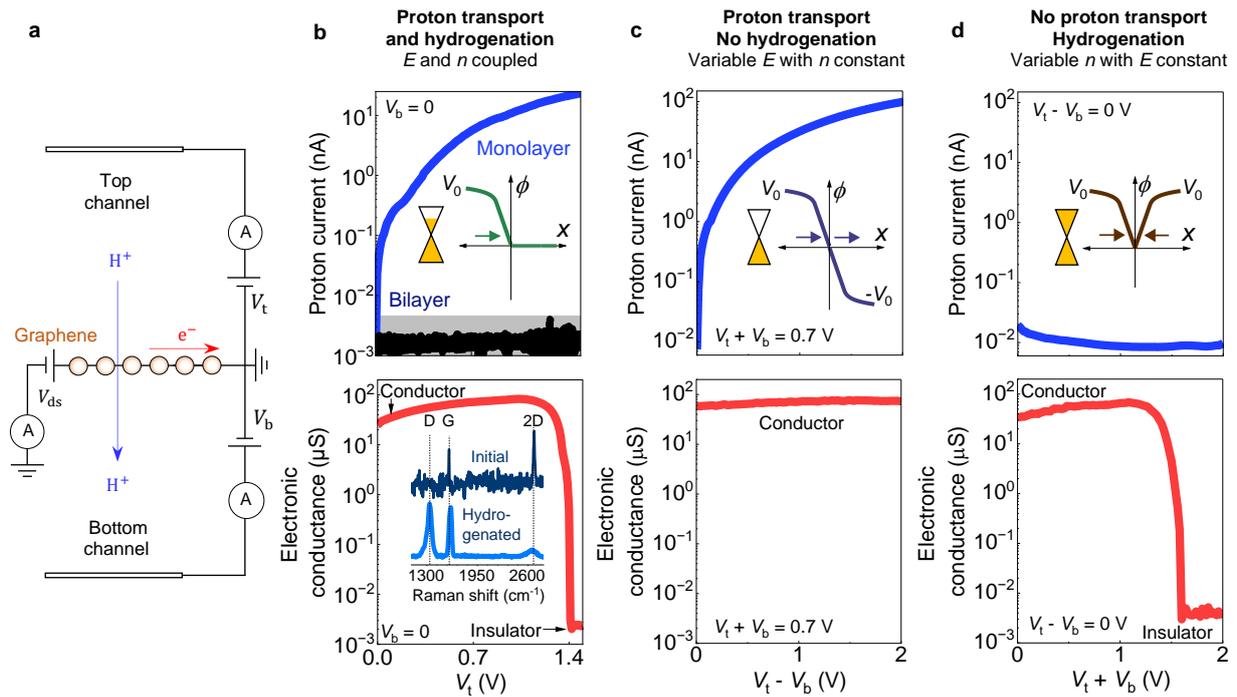

**Figure 1| Selective control of proton transport and hydrogenation in double-gated graphene devices. a**, Schematic of devices in this work. **b**, Top panel, proton current vs voltage characteristics from monolayer graphene devices for $V_b = 0$ (blue curve). In this gate configuration $E$ and $n$ are coupled. Black flat curve, corresponding characteristics for bilayer graphene. Grey area, resolution background determined by parasitic leakage currents. Bottom panel, corresponding in-plane electronic conductance, demonstrating a conductor-insulator transition. $V_{ds}$ = 0.5 mV. Bottom inset, Raman spectra collected in the conductor (dark blue) and insulator regime (light blue, displaying a sharp $D$ band). Background signal from the electrolyte was subtracted from the spectra; spectrum of hydrogenated graphene is divided by factor of 5 for clarity. **c**, Top panel, proton current in devices as a function of $V_t - V_b$ for constant $V_t + V_b$ = 0.7 V. In this gate configuration $n$ is constant and $E$ is variable. Bottom panel, corresponding in-plane electronic conductivity showing that hydrogenation does not take place. **d**, Top panel, proton current in devices as a function of $V_t + V_b$ for constant $V_t - V_b = 0$, showing negligible proton transport. In this gate configuration $E = 0$ and $n$ is variable. Bottom panel, corresponding in-plane electronic conductivity showing that graphene becomes hydrogenated. Top insets in panels b-d, schematics of gate potential (ϕ) vs distance ($x$) from graphene for the different gate configurations ($V_t = V_0$ and $V_b = 0$, green; $V_t = V_0$ and $V_b = -V_0$, purple; $V_t = V_b = V_0$, brown). The gate potentials shift the Fermi level of graphene with respect to the neutrality point (yellow schematics). Horizontal arrows in schematics mark the interfacial electric field induced by each of the gates, which add up to yield the total $E$ in graphene.

**Double-gated graphene devices.** Our device configuration consists of mechanically exfoliated graphene suspended over a micrometre sized hole (10 µm diameter) etched into silicon-nitride substrates, as previously reported[1] (Device fabrication in Methods, Extended Data Fig. 1). The resulting suspended films were coated on both sides with a non-aqueous proton conducting electrolyte with large (>4 V) electrochemical stability window (HTFSI dissolved in PEG[14]) and contacted with two proton injecting electrodes ($PdH_x$). For reference, we also measured devices using electrolytes in which free protons were exchanged for $Li^+$ ions (substituting HTFSI for LiTFSI). The graphene films were then connected into the electrical circuit shown schematically in Fig. 1a. Two sets of (gate) voltages, $V_t$ and $V_b$, are applied between graphene and each of the $PdH_x$ electrodes, which allow controlling the potential on each graphene-electrolyte interface independently (Extended Data Fig. 2). The applied gate voltages drive the proton transport current in the device (Transport measurements in Methods and Extended Data Fig. 3), which is the first process we investigate here. The second process is hydrogenation. To measure the conductor-insulator transition induced by this process, we measure graphene's in-plane electronic conductivity (applying a drain-source voltage, $V_{ds}$) as a function of the applied gate voltages. Hence, this setup enables the simultaneous measurement of the out-of-plane proton and in-plane electronic conductivity of graphene.

**Field-effect enabled selectivity.** To appreciate the advantages of using two gates, we first characterise proton transport and hydrogenation for the case in which one of the gates, the bottom one, is set to zero, $V_b = 0$. Fig. 1b top panel (blue curve) shows that applying a bias to the top gate leads to proton transport through graphene ($V_t > 0$, $V_b = 0$). Reference devices fabricated with bilayer graphene, impermeable to protons[1], displayed no current within our experimental resolution (black curve in Fig. 1b). Moreover, monolayer graphene devices measured with $Li^+$ conducting electrolyte displayed no current either (Extended Data Fig. 4), consistent with graphene's known impermeability to all ions[3,4]. These experiments confirm that the transport current observed in monolayer devices is indeed due to proton transport. On the other hand, Fig. 1b bottom panel shows that as we drive proton transport through monolayer graphene, the in-plane electronic conductivity drops by 4 orders of magnitude around $V_t \approx 1.4$ V – turning graphene into an electronic insulator – and that a prominent $D$-band appears in graphene's Raman spectrum (Fig. 1b, bottom inset and Extended Data Fig. 5). These results therefore show that if one of the gates is set to zero, accelerating proton transport with the other gate eventually leads to hydrogenation of the lattice. Unexpectedly, we find that using two gates allows driving strong proton transport through graphene without hydrogenation. Fig. 1c shows the response of the devices when the gate voltages are set such that their sum is fixed, $V_t + V_b = 0.7$ V, but their difference is variable, $V_t - V_b > 0$. This yields strong proton transport current (top panel), but crucially, graphene remains electrically conductive even for large gate potentials, demonstrating that hydrogenation does not take place (bottom panel). The converse is also possible. Fig. 1d shows that setting the gate potentials such that their difference is fixed, $V_t - V_b = 0$, but their sum is variable $V_t + V_b > 0$, can suppress proton transport (top panel), but now graphene becomes hydrogenated (bottom panel). These results demonstrate that double-gated devices allow driving the two processes selectively, even at high bias, which is not possible using only one gate.

To understand why double gating enables this decoupling, we recall the well-established fact[25-30] that in double-gated 2D crystals $E \propto V_t - V_b$, whereas $n$ depends only on $V_t + V_b$. This point is discussed quantitatively in Methods but can be understood qualitatively as follows. Consider the case in which both gates are fixed at the same potential ($V_t = V_b = V_0$), illustrated in Fig. 1d inset. Both gates shift graphene's Fermi level ($\mu_e$) in the same direction, and since $n \propto \mu_e^2$, this raises $n$. However, the electric fields in the two graphene-electrolyte interfaces – the gradients of the gate potentials (horizontal arrows, Fig. 1d inset) – point in opposite directions, so they yield zero total $E$ in the 2D crystal. In this case then, $V_t + V_b = 2V_0$ and $V_t - V_b = 0$, which illustrates that $n$ and $E$ are determined by sum and the difference of the gate potentials, respectively. Conversely, if the gates have opposite polarity ($V_b = -V_0 = -V_t$, Fig. 1c inset), $\mu_e$ is driven in opposite directions by the gates. This yields zero induced $n$ but the electric fields induced by each gate now point in the same direction, yielding high $E$ in graphene ($V_t + V_b = 0$, $V_t - V_b = 2V_0$). The solution to the corresponding electrostatic equations shows that $E$ and $n$ are indeed functions of $V_t - V_b$ and $V_t + V_b$, respectively, and depend only on the capacitance of the electrolyte and fundamental constants (Estimation of $E$ and $n$ in Methods). Direct characterisation of such capacitance ($C \approx 20$ μF cm$^{-2}$, Extended Data Fig. 6) reveals that our devices can achieve $E \sim$ V nm$^{-1}$ and $n \sim 10^{14}$ cm$^{-2}$. This discussion therefore reveals that decoupling proton transport from hydrogenation is possible here due to the independent control of $E$ and $n$ in double-gated graphene. We explore this decoupling systematically by mapping both processes in terms of these variables.

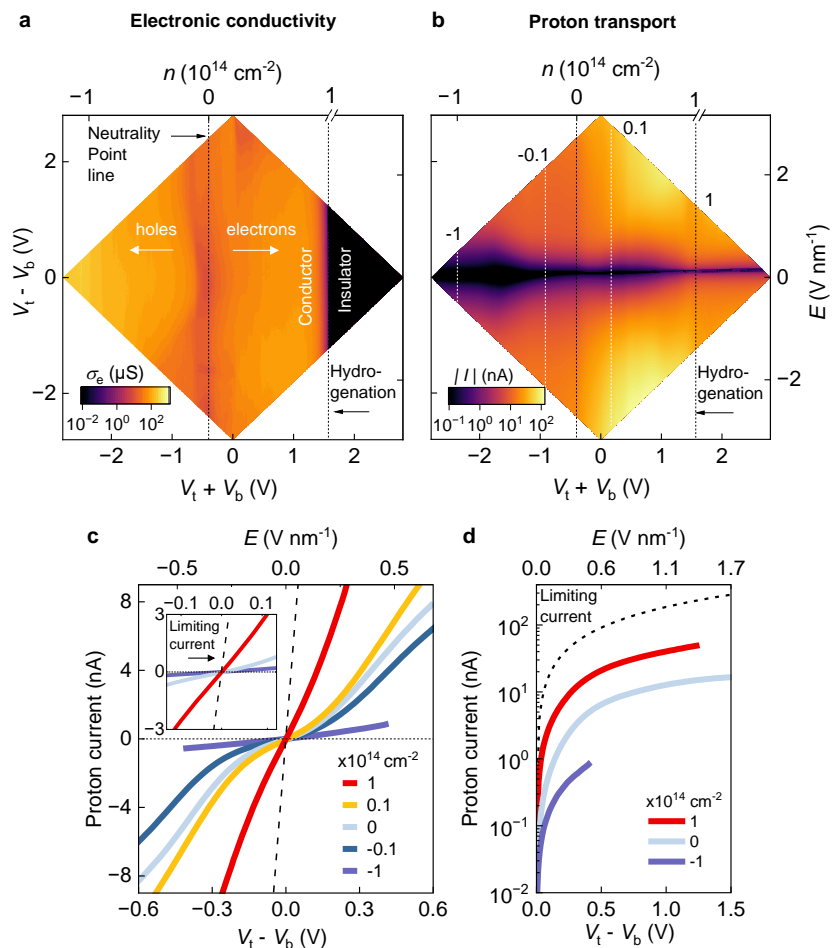

**Figure 2| Proton and electronic transport with independent control of $E$ and $n$ in double-gated graphene. a**, Map of in-plane electronic conductance, $\sigma_e$, as a function of $E$ and $n$ (left and bottom axes show that these variables are controlled by $V_t - V_b$ and $V_t + V_b$, respectively). Top axis ($n$) is cut off after the hydrogenation transition because the scale no longer applies (Estimation of $E$ and $n$ in Methods). The neutrality point line is visible as a slightly darker band in the map (marked with a vertical dashed line). The conductor-insulator transition due to hydrogenation is apparent as a sharp boundary for high electron doping (marked with dashed line). **b**, Map of proton transport current, $I$, as a function of $E$ and $n$. Dashed lines mark cross sections at constant $n$ illustrated in panel **a**. The number labels mark the fixed $n$ for each cross section in units of $10^{14}$ cm$^{-2}$. **c**, Proton transport current

as a function of *E* for constant *n*. Each *I-E* curve is a cross section taken from panel **b**. The number labels mark the fixed *n* for each cross section in units of $10^{14}$ cm$^{-2}$. Dashed line marks the limiting current enabled by the electrolyte obtained from devices without graphene (Extended Data Fig. 9). Dotted horizontal line, guide to the eye. Inset, zoom in from main panel for *E* < 0.1 V nm$^{-1}$, in which the transport characteristics are linear. **d**, Proton transport current as a function of *E* for constant *n* in the high *E* regime. Number labels mark the fixed *n* for each cross section in units of $10^{14}$ cm$^{-2}$.

**Proton and electronic transport maps**. We start with hydrogenation. Graphene's in-plane electronic conductivity was mapped by sweeping *E* for a fixed *n* and then stepping *n* from hole- towards electron-doped regions. Fig. 2a and Extended Data Fig. 5 show that the conductivity displays a local minimum, the so-called charge neutrality point (NP), visible as a vertical band at -0.35 ± 0.15 V that splits the map between hole- and electron-doped regions. To the right of the NP, where graphene is electron-doped, we find a conductor-insulator transition evident as a sharp boundary at $n \approx 1 \times 10^{14}$ cm$^{-2}$ that is accompanied by the sudden appearance of a *D* band in graphene's Raman spectrum (Extended Data Fig. 5). This shows that graphene is hydrogenated, and we find that this state is retained unless a negative gate voltage is applied, resulting in a hysteretic dependence of the process on gate voltage (Extended Data Fig. 5). The density of adsorbed H atoms in hydrogenated samples can be estimated directly from the intensity of the *D* band in their Raman spectrum as ~$10^{14}$ cm$^{-2}$ (Raman spectroscopy in Methods), consistent with graphene's electron doping at the hydrogenation transition. No *D* band was observed in reference devices measured with Li$^+$ conducting electrolyte at any applied bias (Extended Data Fig. 4), confirming that the observed phenomena are indeed due to proton adsorption in graphene. The found dependence of the hydrogenation process on *n* can be rationalised using a classical analytical model. This shows that the energy barrier for hydrogenation is effectively suppressed for the potential configuration that leads to large electron doping (Extended Data Fig. 7), in agreement with our density functional theory calculations (Extended Data Fig. 8, see also Hydrogenation transition in Methods). The dependence can also be understood by noticing that *n* is related to the electrochemical potential of electrons with respect to the NP in graphene as $\mu_e \propto \sqrt{n}$. Such relation implies that the hydrogenation process is driven by $\mu_e$, consistent with the well-established notion that electrochemical charge transfer processes are driven by this potential (Electrochemical description of the hydrogenation process in Methods).

We now discuss proton transport. Fig. 2b shows the proton transport map obtained simultaneously with the electronic map in Fig. 2a. To analyse it, we take cross sections of the map at constant *n* both from hole- and electron-doped regions (dotted lines in Fig. 2b). Fig. 2c shows that proton transport is driven by *E*, but the current is notably larger when graphene is electron doped. This finding can be analysed quantitatively in the low *E* regime (*E* < 0.1 V nm$^{-1}$), where the proton transport characteristics are linear. Fig. 2c inset shows that for electron doping of ≈ 1 x $10^{14}$ cm$^{-2}$, graphene is ≈ 30 times more proton conductive than for hole doping with the same concentration and reaches values just ~5 times lower than the limiting conductivity enabled by the electrolyte (black dashed line in Fig. 2c). For larger fields, the transport characteristics become non-linear and for even larger fields of > 0.5 V nm$^{-1}$, the transport for all doping regimes trace the limiting current enabled by the electrolyte, but attenuated according to their doping (Fig. 2d). This field effect is reversible and does not arise from changes in the electrolyte conductivity with bias (Extended Data Fig. 9). It is an intrinsic effect of graphene at high *E* and *n*. The dependence of proton transport on these variables can be understood using a similar analytical model to the one used to study the hydrogenation transition. The model reveals that configuring the gate potentials to dope graphene with electrons distorts the potential energy profile for the incoming protons, facilitating their transport with respect to the case where graphene is not doped. Large *E* also distorts this energy profile, providing incoming protons with energy comparable to the barrier height, thus facilitating the transport for any doping configuration (Extended Data Fig. 7). This model is consistent with our DFT calculations, which show that *E* reduces the energy barrier for proton transport with respect to the zero-field case (Extended Data Fig. 10) and as such should result in a strong acceleration of the transport.

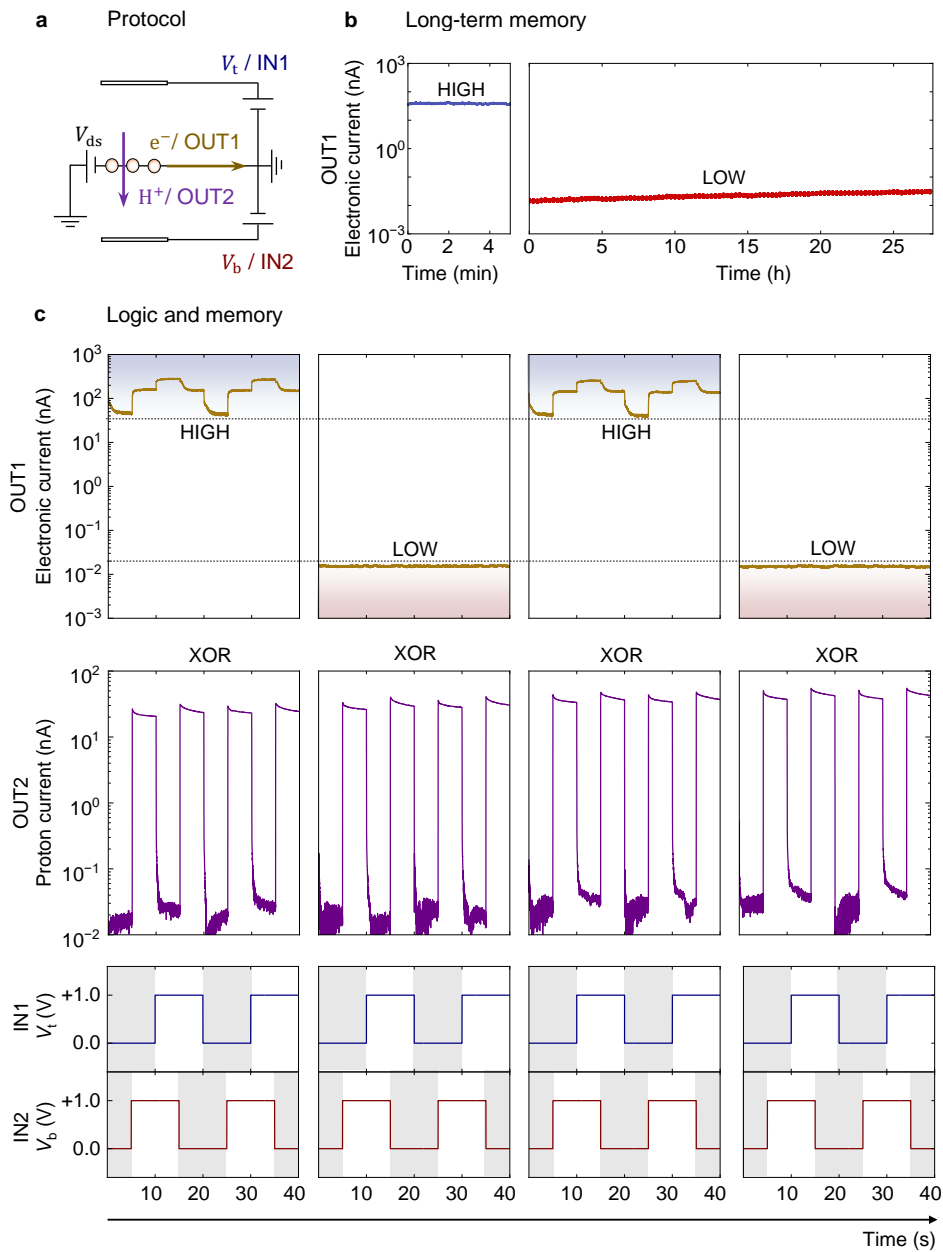

**Figure 3| Robust and precise switching of proton transport and hydrogenation in double-gated graphene enables proton-based logic-and-memory devices**. **a**, Schematic of signal protocol. $V_t$ and $V_b$ are set as the logic inputs, IN1 and IN2, respectively. The electronic current is defined as OUT1 and functions as the memory state. Proton current is set as the logic output OUT2. **b**, Demonstration of the long-term memory state retention. Left, the device is in the HIGH (conductive) state whose retention is tested under IN1 = IN2 = 0. Right, the device is pre-programmed into the LOW (insulating) state whose retention is also tested under IN1 = IN2 = 0 V. The LOW state is programmed applying 1.5 V on both gates and the HIGH state is recovered applying -1.4 V on both gates. **c**, Demonstration of

simultaneous logic and memory operations using the out-of-plane proton system as logic gate and the in-plane electronic system as memory unit. The bottom two panel rows show the input signals, IN1 and IN2, which are squared waveforms with various combinations of input values (00, 01, 11, 10, in two cycles). The middle panel row shows the dynamic response of the proton currents. Under the input levels (00 or 11), OUT2 displays low currents of ≈ 20 pA; whereas for different levels (01 or 10), it displays high currents of ≈ 30 nA, thus demonstrating an XOR logic gate. The top panel row shows the dynamic response of the electronic current (source-drain bias 0.5 mV), measured simultaneously with the out-of-plane proton current. Graphene's in-plane electronic system was pre-programmed into dehydrogenated HIGH memory state and hydrogenated LOW memory state (marked with dashed lines and shaded areas in the graphs).

**Precise and robust control of processes.** We investigate the precision and robustness with which we can selectively drive the proton transport and hydrogenation processes. To that end, we evaluate the devices' performance in logic and memory applications (Fig. 3a). The in-plane electronic system is used as memory unit and the hydrogenation process is exploited to programme two memory states, HIGH (conducting) and LOW (insulating), such that the ratio of their conductivity exceeds $10^3$. Fig. 3b shows that the memory states are non-volatile and are retained for over one day – as long as we decided to measure. This non-volatility is a consequence of the hysteretic dependence of the hydrogenation process on $n$ discussed above (Extended Data Fig. 5). With the electronic system pre-programmed, we use the out-of-plane proton transport current to perform logic operations. The two gates are used to apply high $E$ to drive proton currents and perform logic operations, but with $n$ that does not disturb the pre-programmed memory state in the electronic system. Fig. 3c shows that the proton transport system yields an XOR logic operator with on-off ratios of over 3 orders of magnitude and that during its operation the in-plane electronic memory state remains in its pre-programmed (conducting or insulating) state. This demonstrates that the control of the processes is robust and precise enough to enable computation applications. Here, graphene performs both logic and memory functions via independent control of its proton and electronic transport properties. This combination of functions in the same physical area of the device would eliminate[34] the need of peripheral circuits between logic and memory components in prospective arrays of these devices, making them more energy efficient and compact. 3D ensembles that exploit the independent proton and electron current pathways could enable even denser proton-based logic-and-memory networks[35].

**Outlook.** We showed that double-gated graphene devices enable precise and robust control of proton transport and hydrogenation via independent control of $E$ and $n$ in graphene. We showed that in double-gated graphene, the proton current itself can be used to perform logic operations, of interest in the field of electrochemically gated electronic materials[36,37]. We showed that field effects strongly accelerate proton transport, important to its applications in proton conducting membranes[1,2], catalysis[9,10] or isotope separation[11,12]. More generally, we show that double-gated 2D crystals enable mapping processes in electrode-electrolyte interfaces as a function of $E$ and $n$, which cannot currently be achieved without double gating. We call these maps electrochemical charge-field maps (ECFM) to differentiate them from conventional double-gating, in which the gate (electrolytic) current is not monitored as a meaningful variable. The selective acceleration of proton transport and hydrogenation reported here suggests that similar 2D crystal devices could selectively drive other coupled interfacial processes. Given recent advances in embedding catalysts into 2D crystals[38], such processes could in the future perhaps include reactions like $CO_2$ reduction, which compete with the much faster hydrogen evolution reaction[39]. Fundamentally, our work expands the parameter space over which electrochemical processes in 2D crystals can be studied.

**Methods**

**Device fabrication.** Apertures 10 μm in diameter were etched in silicon-nitride substrates (500 nm $SiN_x$) using photolithography, wet etching, and reactive ion etching, as previously reported[1]. Source and drain electrodes (Au/Cr) were patterned using photolithography and electron beam evaporation. Mechanically exfoliated mono- and bilayer graphene crystals were transferred over the apertures and on the electrodes (Extended Data Fig. 1a). We selected crystal flakes with a rectangular shape, with their long side several tens of micrometres, to form a conducting channel between the source and drain electrodes. The flake's width is chosen to be only a couple micrometres wider than the aperture (Extended Data Fig. 1b), which ensures that a whole cross-section of flake can be gated with the two gates effectively. Such cross-section of the flake becomes the active area of the device, with the non-gated areas acting as electrical contacts to the gated section. An SU-8 photo-curable epoxy washer with a 15 μm-diameter hole was transferred over the flake and on the source and drain electrodes[3] with the hole in the washer aligned with the aperture in the silicon-nitride substrate (Extended Data Fig. 1b). The polymer seal ensures that the electrodes are electrically insulated from the electrolyte. The electrolyte used was 0.18 M bis(trifluoromethane)sulfonimide (HTFSI) dissolved in poly(ethylene glycol) (average molecular weight Mn of 600)[14] never exposed to ambient conditions. This electrolyte was drop-cast on both sides of the device in a glovebox containing inert gas atmosphere. For reference, we measured devices prepared in the same way, except HTFSI was substituted for LiTFSI in the electrolyte. Palladium hydride foils (~0.5 $cm^2$) were used as gate electrodes. The device was placed inside a gas-tight Linkam chamber (HFS600E-PB4) filled with argon for electrical measurements.

**Transport measurements.** For electrical measurements, a dual-channel Keithley 2636B sourcemeter was used to bias both gates. The applied voltages yield two proton current signals, top ($I_t$) and bottom ($I_b$) channel current, also recorded with the Keithley sourcemeter. These two signals quantify the two halves of the proton transport circuit: proton transport from one gate electrode towards graphene and then from graphene towards the other gate electrode. Due to graphene's proton permeability these two currents are effectively identical (Extended Data Fig. 3), differing by only ~1%. For this reason, it is sufficient to use only one of them to unambiguously characterise proton transport in the device, $I$. A second Keithley 2614A sourcemeter was used both to apply drain-source bias ($V_{ds}$) and to measure the electronic resistance ($\sigma_e$).

To confirm that the two gates are independent, we contacted the electrolyte in the top channel with a reference electrode (PdHx with the same size of the gates), whose potential, $V_t^{ref}$, was monitored with a Keithley 2182A Nanovoltmeter (Extended Data Fig. 2a). In the first experiment, we swept $V_b$ for various fixed $V_t$. Extended Data Fig. 2b shows that $V_t^{ref} = V_t$ for all values of $V_b$, within the experimental scatter of <4 mV. This demonstrates that sweeping $V_b$ does not affect $V_t$. In the second experiment, we swept $V_t$ for various fixed $V_b$. This measurement also showed that $V_t^{ref} = V_t$ (Extended Data Fig. 2c), which demonstrates that a fixed $V_b$ does not affect $V_t$ either. These experiments therefore demonstrate that the gates are independent from each other.

To obtain maps of the proton and electronic systems, we measured $I$ and $\sigma_e$ simultaneously as a function of $V_t$ and $V_b$. The maps were obtained using software that allowed controlling $V_t - V_b$ and $V_t + V_b$ as independent variables. We swept $V_t - V_b$ (for a fixed $V_t + V_b$) at a rate of 10 mV $s^{-1}$ for each gate and stepped $V_t + V_b$ with 10 mV intervals. The maximum $V_t$ or $V_b$ applied was ± 1.4 V, which resulted in a maximum $V_t + V_b$ and $V_t - V_b$ of ± 2.8 V. We normally did not apply *V*-bias beyond these voltage ranges to avoid damaging the devices.

**Raman spectroscopy.** For Raman measurements, the graphene devices were left in the same gas-tight Linkam chamber (HFS600E-PB4) used for electrical measurements, which has an optical window. The Raman spectra of devices were measured as a function of applied *V*-bias using a 514 nm laser. The background signal from the electrolyte was removed for clarity, which results in relatively weak Raman spectra for pristine graphene (Extended Data Fig. 5). After hydrogenation, a strong *D* peak appears and the intensity of the *G* peak increases while the 2D peak becomes broader, in agreement with previous work[14]. The density of adsorbed hydrogen atoms in hydrogenated graphene was estimated from the ratio of peak-height intensities of the *D* and *G* band[40,41], $I_D/I_G$. In our devices $I_D/I_G \approx 1$, which corresponds to a distance between H atoms in graphene of $L_D \approx 1$ nm. An equivalent analysis using the integrated-area ratio in these peaks[42], $A_D/A_G \approx 2$, yields $L_D < 1.2$ nm. Both estimates yield a density of H atoms of $\approx 1 \times 10^{14}$ cm$^{-2}$, in agreement with previous reports on hydrogenated graphene[13,14].

**Electrolyte characterisation.** To characterise the limiting conductivity of our devices, we measured devices similar to those described above but in which the aperture in the silicon nitride substrate was not covered with graphene ('open hole device'). Extended Data Fig. 9 shows that the *I-V* characteristics of these open hole devices were linear within all the *V*-bias range used in this work. This demonstrates that the field effect we observe in graphene devices does not arise from changes in the electrolyte conductivity at high *V*, consistent with the known large electrochemical window of this electrolyte (4-5 V[43]).

To characterise the capacitance of the electrolyte, we patterned two Au electrodes on a silicon nitride substrate using photolithography and electron beam evaporation. The electrodes were connected into an electrical circuit and a polymer mask was used to cover all the electrodes except for an active area that was exposed to the electrolyte. The electrodes' area differs by a factor of ~50, which ensures that the total capacitance is dominated by the smaller one and allows observing differences, where present, of the devices response under positive and negative potentials[44]. Cyclic voltammetry (CV) measurements with scan speeds ranging from 1 to 40 mV s$^{-1}$ were performed over the voltage range of -0.1 V and 0.1 V. Extended Data Fig. 6a shows that the CV curves display no redox peaks nor asymmetry between the positive and negative voltage branches. The area normalised capacitance of the electrolyte, *C*, can then be obtained from the CV curves from the expression[44,45]: $C = (A \Delta V v)^{-1} \int I \, dV$, where *A* is the active area of the electrode, $\Delta V$ is the voltage range in the CV, *I* is the current and *v* is the scan speed. Extended Data Fig. 6b shows the extracted *C* as a function of *v*. For the smallest scan rate (1 mV s$^{-1}$), we find $C \approx 30$ μF cm$^{-2}$. This value decreases with *v*, as expected. Since in our measurements we use *v* = 10 mV s$^{-1}$, we use the value obtained at such *v*, $C \approx 20$ μF cm$^{-2}$, in our estimates involving *C*.

**Estimation of *E* and *n*.** The Debye length in our electrolyte (0.18 M salt and solvent dielectric constant $\varepsilon_r \approx 10$)[46,47] can be estimated as $\lambda_D \approx 0.3$ nm. Given this and the relatively large gate bias used in this work, the electrical potential across the graphene-electrolyte interface in our devices drops almost entirely across the Stern layer. Hence, each of the gate potentials can be described using a parallel plate capacitor model, which we use to derive the relations between the gate potentials and *E* and *n*, as shown in refs. [25-27,48,49].

To derive the relation between $V_t + V_b$ and $n$, we note that if only one gate (top) operates, the charge induced is $neC^{-1} = (V_t - V_t^{NP})$, where the superscript NP marks the neutrality point and $e$ is the elementary charge constant. An equivalent relation holds for the top gate. Hence, the total charge from both gates is given by the addition of their contributions: $neC^{-1} = (V_t + V_b) - \Delta^{NP}$, where $\Delta^{NP} \equiv V_t^{NP} + V_b^{NP}$. To consider the quantum capacitance of graphene, we note that $\mu_e = \hbar v_F \sqrt{\pi n}$, with $v_F \approx 1 \times 10^6$ m s$^{-1}$ the Fermi velocity in graphene and $\hbar$ the reduced Planck constant. This changes the relation to[50]:

$$(V_t + V_b) - \Delta^{NP} = neC^{-1} + \hbar v_F e^{-1}(\pi n)^{1/2} \quad (1)$$

From the estimate of $C$ above, we get $(V_t + V_b) - \Delta^{NP} \approx 0.8 \times 10^{-14}$ V cm$^2$ $n$ + $1.16 \times 10^{-7}$ V cm $n^{1/2}$. Note that this description is accurate only if the Fermi energy of the system is outside a bandgap[31]. Hence, we only use it when graphene is conductive. After the hydrogenation transition, we cannot assess $n$, as indicated with breaks in the top axes in Fig. 2.

To derive the relation between $V_t - V_b$ and $E$, we note that if only one gate (top) operates, the electric field induced by the gate is $E_t = en_t (2\varepsilon)^{-1} = C (2\varepsilon)^{-1} (V_t - V_t^{NP})$, with $\varepsilon$ the dielectric constant of the solvent. Note that the electric field points in the direction between graphene and its corresponding electrical double layer, which we define as $+x$ for the top gate. An equivalent relation holds for the bottom gate, except $E_b$ points in the $-x$ direction (towards its corresponding electrical double layer). The total electric field in graphene is then:

$$E = E_t - E_b = C (2\varepsilon)^{-1} (V_t - V_b) \quad (2)$$

This yields $E \approx 1.13 \times 10^9$ m$^{-1}$ $(V_t - V_b)$.

**Analytical model of proton transport and hydrogenation in double-gated graphene.** We used an analytical model to illustrate how the gate voltages affect proton transport and hydrogenation in double gated graphene. The energy barrier for proton transport through the centre of the hexagonal ring in graphene is modelled using a Gaussian function: $V_p = G_0 \exp(-x/W)^2$, with $G_0 = 0.8$ eV the barrier height determined experimentally in the low electric field limit[1] and $W = 0.5$ Å the barrier width (Extended Data Fig. 7b). For the hydrogenation process, the proton is directed towards the top of a carbon atom in graphene.

The potential energy profile for the hydrogenation process consists of two parts, the energy barrier ($V_{Hb}$) and the adsorption well ($V_{Ha}$). The energy barrier is modelled with a Lorentzian-type function: $V_{Hb} = V_0 [((x-|x_0|)/d_0)^3 + 1]^{-1}$, with $V_0 = 0.2$ eV the barrier height, $x_0 = 1.7$ Å the distance between the barrier and graphene and $d_0 = 0.4$ Å the barrier width. The 3$^{rd}$ power in the denominator models long-range van der Waals interactions. The adsorption well is modelled with a Lorentzian: $V_{Ha} = V_1 [((x-|x_1|)/d_1)^2 + 1]^{-1}$, with $V_1 = -0.8$ eV the well depth, $x_1 = 1.1$ Å the distance between the well and graphene and $d_1 = 0.25$ Å the well's width. Note that the well is modelled to be strongly repulsive at $x = 0$ to capture the repulsion between C and H atoms at very short distances. The parameters for these functions are taken from DFT calculations[14,51] and the total potential for the hydrogenation process is then $V_{Hb} + V_{Ha}$ (Extended Data Fig. 7a).

The gate potential profiles ($V_t$ and $V_b$) are modelled with a Guoy-Chapman-Stern model[24], using dielectric constant $\varepsilon_r = 10$ for the solvent, electrolyte concentration of 0.18 M and Stern layer thickness of 0.4 nm. The resulting gate potentials drop almost exclusively over the Stern layer (Extended Data Fig. 7) and, hence, the graphene-electrolyte interface behaves as a capacitor, as discussed above. We find that the qualitative findings of our model are relatively insensitive to the specific parameters of the GCS model if the Stern layer is >0.3 nm. The superposition of the potentials for each of the processes with the gate potentials model the behaviour of the devices.

To illustrate the role of the gates in the hydrogenation process, we set them to yield $n = 1.2 \times 10^{14}$ cm$^{-2}$ but $E = 0$. Extended Data Fig. 7a shows that this distorts the potential energy profile for hydrogenation, such that the hydrogenation barrier is now easily surmounted by incoming protons, which become trapped in the adsorption well and hydrogenate graphene. To illustrate the role of $E$ in the proton transport process, we set the gates to produce large $E = 1.7$ V nm$^{-1}$ but $n = 0$ cm$^{-2}$. Extended Data Fig. 7b shows that this distorts the potential energy profile for proton transport, such that the barrier is now easily surmounted by a proton moving in the direction of the electric field (from the $-x$ to the $+x$). To illustrate the role of electron doping in proton transport, we set the gates to give large $n = 1 \times 10^{14}$ cm$^{-2}$ but $E = 0.67$ V nm$^{-1}$. Extended Data Fig. 7c shows that this also distorts the potential energy profile for the incoming protons, resulting in facilitated transmission over the barrier. The model illustrates that the distortion of the energy profile for incoming protons due to $E$ and $n$ in these devices is comparable to barrier height itself. For this reason, these variables dominate the devices' response and previously identified effects, like strain and curvature, should play a secondary role.

**Electrochemical description of the hydrogenation process.** The transport data in the manuscript are described using the variables $E$ and $n$. However, it is equivalent to describe the system using the electrochemical potential of electrons in graphene with respect to the NP, $\mu_e$, instead of $n$. Indeed, one important property of graphene is that $n$ and $\mu_e$ are related via the formula $\mu_e = \hbar v_F \sqrt{\pi n}$, with $v_F \approx 1 \times 10^6$ m s$^{-1}$ the Fermi velocity in graphene and $\hbar$ the reduced Planck constant. This relation is fundamental, arising from the density of states in the material and holds exactly in experimental systems[52,53]. Moreover, this relation is valid independently of whether the material is gated or not. Hence, the top $x$-axis in the hydrogenation map in Fig. 2a can be re-expressed in terms of $\mu_e$, which illustrates that the hydrogenation process is driven by $\mu_e$. This is consistent with the well-established notion that electrochemical charge transfer processes are driven by this variable. Note that while the relation between $n$ and $\mu_e$ is fixed, applying a gate voltage to graphene, shifts both variables[31,54]. However, these variables are not independent, as discussed above. To determine their dependence on the gate voltage, it is necessary to establish the electrostatic gate capacitance, $C$ (see Extended Data Fig. 6). Once $C$ is determined, the dependence of $n$ (or $\mu_e$) on gate voltage is described by Eq. (1) in Methods.

**Hydrogenation transition.** It is instructive to compare our results with previous work on plasma-hydrogenated graphene[13,55]. The latter samples typically display ~100 times higher electronic resistivity than in their non-hydrogenated state. In contrast, in both our work and in ref. [14] this factor is of about 10$^4$, yielding an insulating state mostly insensitive to the gate voltage. There are at least two possibilities for this difference. The first one is that the H atom densities obtained by the different methods are different. Indeed, while the Raman spectra of both this work and ref. [13] yield $I_D/I_G \approx 1$, this could arise both from a H atom density of <10$^{12}$ cm$^{-2}$ or ~10$^{14}$ cm$^{-2}$, because of the bell-shape[40,41] of the $I_D/I_G$ vs defect-density dependence. To decide which one applies, it is therefore necessary to look for further evidence of disorder in the spectra. The Raman spectra in ref. [13] displayed a sharp *2D* band, typical of ordered samples, which suggests that the H density was in the regime of <10$^{12}$ cm$^{-2}$. This contrasts with the *2D* band in our spectra, which is smeared, consistent with ~10$^{14}$ cm$^{-2}$. The second possibility is that both systems have the same H density. In this case, the higher resistivity could arise from a more disordered H atom distribution. Indeed, H atoms in plasma hydrogenated graphene are known to cluster[33,56], which reduces the number of effective scattering centres proportionally to the number of atoms in the cluster. The reduction could be significant because the scattering radius around each H atom extends to second neighbours (9 carbon atoms)[33,56,57]. The electrochemical system could be less prone to clusters, for example, because the electrolyte stabilises the proton as it adsorbs on graphene, making the reaction more likely to happen than in vacuum, thus yielding a more random distribution.

Another difference between the two hydrogenation methods is their reversibility. Ref. [13] showed that the plasma hydrogenation process can be almost completely reversed by annealing the material in Ar atmosphere. However, a *D* band was still notable after annealing and some of graphene's electronic properties were not fully recovered[13]. This imperfect reversibility was attributed[13] to the presence of vacancy defects introduced during the plasma exposure. In both our work and in ref. [14] the hydrogenated transition is fully reversible, with no *D* peak apparent in the Raman spectra of de-hydrogenated samples.

Another difference with plasma hydrogenated samples is that electrochemical hydrogenation allows studying the dependence of the transition on *n*. This has revealed that the transition is sharp. We attribute this sharpness to a percolation-type transition[58] triggered both by the high density of adsorbed H atoms in the samples and to the carrier scattering associated with them[59,60]. We propose that the insulating state in the samples is therefore a consequence of their high disorder, as suggested in previous works[59,60], rather than a bandgap. This is consistent with experimental studies reporting that a bandgap in plasma hydrogenated samples typically require either patterned distribution of H atoms[61] or much higher H atom density[62] than in the samples in this work.

**DFT calculations of graphene hydrogenation.** Simulations of graphene hydrogenation were performed using the Vienna ab initio Simulation (VASP) Package[63,64,65,66]. Electron-ion interactions were modelled using the projector augmented wave method and the exchange correlations of electrons were modelled with the Perdew-Burke-Ernzerhof (PBE) type generalised gradient approximation (GGA) functional[67]. Spin polarisation was considered and the van der Waals interactions were incorporated by using the Grimme's DFT-D3 method[68]. Initial crystal structure relaxation was performed with a force criterion of 0.005 eV/Å and an electronic convergence of $10^{-6}$ eV, accelerated with a Gaussian smearing of 0.05 eV. The energy cut-off was set at 500 eV, and Monkhorst-Pack k-point mesh with a reciprocal spacing of $2\pi \times 0.025/\text{Å}$ was implemented, which ensured energy convergence to 1 meV. We constructed a cubic simulation, consisting of a 4x7 orthogonal supercell with 112 carbon atoms placed at the centre in the *z* direction (*i.e.*, perpendicular to the 2D plane), and with a vacuum slab to prevent interactions between adjacent periodic images. After relaxation, the energy barriers for a proton to be adsorbed on top of a carbon atom, under vacuum conditions, were then calculated by ab initio molecular dynamics (AIMD) simulations using the microcanonical NVE ensemble and the same convergence criteria mentioned above. We used a time step of 0.1 fs and a minimal initial kinetic energy for the proton in the direction perpendicular to the 2D layer, as previously reported[1,69]. A dipole correction was implemented to study the influence of an external electric field perpendicular to the 2D layer (along the *z* direction)[70]. Due to the periodic boundary conditions, this dipole is repeatedly inserted in all the simulation boxes along the z direction, yielding a constant electric field in the direction perpendicular to graphene[70].

Extended Data Fig. 8 shows the calculated potential energy curves for the proton-graphene system. The curves are calculated as a function of distance between the proton and the top of a C atom site with a fully relaxed lattice. The potential energy curves display a minimum (adsorption well) at ≈1.14 Å (C-H bond) and a small adsorption barrier around ≈2 Å, in agreement with previous studies[14]. We find the electric field distorts the potential energy profile for hydrogenation, favouring the process in agreement with the analytical model. For reference, we also performed calculations using the non-local optB88-vdW and the hybrid functional HSE06. These resulted only in minor differences (<0.1 eV) in the hydrogenation barrier height compared to PBE.

**DFT calculations of proton transport through graphene.** The DFT calculations of proton transport through graphene were performed using the Vienna ab initio Simulation Package (VASP)[63,64,65,66] and PWSCF with Quantum Espresso (QE). We used the optB88-vdW[71] functional, with a 3 x 3 x 3 Γ-centred k-point grid, a 1000 eV energy cut-off with hard pseudopotentials[72,73], and a force convergence criterion of 0.03 eV/ Å. We used a $4 \times 4$ unit cell with a vacuum separating periodically repeating graphene sheets of 12 Å for pristine graphene and ~23 Å for hydrogenated graphene. The zero electric field energy profiles were computed using the climbing image nudged elastic band (cNEB)[74] with VASP. Charged cells were employed to describe the protons in the simulations with a uniform compensating background. In the model, protons transfer from a water molecule on one side of graphene to another one on the opposite side. Using these two water molecules minimises spurious charge transfer from the graphene sheet to the proton, as confirmed with a Bader[75] charge analysis. To incorporate the electric field, we modelled the system using QE. Here, we used the optB88-vdW functional[71,76-79], a 3 x 3 x 3 Γ-centred k-point grid, and a 600 Ry energy cut-off. We confirmed that the VASP zero electric field energy barriers are reproduced within ~15 meV in QE. The electric field in QE is simulated as a saw-like potential added to the ionic potential, together with a dipole correction implemented according to ref.[80]. The saw-like potential increases in the region from 0.1 $a_3$ to 0.9 $a_3$, $a_3$ being the lattice vector perpendicular to the graphene sheet, which is placed at the centre of the cell (0.5 $a_3$), then decreases to 0 at $a_3$ and 0. The discontinuity of the sawtooth potential was placed in the vacuum region. The electric field was applied in the perpendicular direction to the graphene basal plane ($z$ direction). For reference, we performed calculations using the PBE-D3 functional, which gave comparable results.

We first calculated the energy profile for proton transport through graphene in the absence of an electric field and for two different levels of H atom coverage of the lattice (0% and 20%). The choice of 20% hydrogenation is to take into account the fact that adsorbed hydrogen atoms typically form dimer structures consisting of two H atoms per 8 carbon atom sublattice[33,56,81], which correspond to a local lattice coverage of ~25%. In agreement with ref.[18], we observe that the energy barrier for pristine graphene reduces by ~30% for 20% H atom coverage. The barrier at zero field we find, $\Gamma_0 \approx$ 3.1-3.4 eV for the different functionals, is larger than the typically found values[7] $\Gamma_0 \approx$ 1-2 eV because in our approach the computed proton trajectory involves a chemisorption state, as described previously[18]. However, we note that the absolute value of the barriers in these simplified models are not especially informative, as discussed in ref.[69]. These models only aim to provide qualitative insights into the influence of $E$ and hydrogenation in proton transport through graphene. Next, we computed the energy profiles along the same pathway used in the zero $E$ calculations, but now including a perpendicular electric field, $E$, along the direction of motion of the proton. Extended Data Fig. 10 shows the energy profiles along the reaction path for the two different levels of hydrogenation of the lattice for various electric fields. Regardless of the extent of hydrogenation, we observe a roughly linear barrier reduction when the electric field is switched on, achieving a ~20% reduction with $E$ ~1 V nm$^{-1}$.

**Logic and memory measurements.** For logic and memory measurements, we defined $V_t$ and $V_b$ as the IN1 and IN2 signals, respectively; and, guided by the maps of the devices, we systematically explored their proton and electronic response under different input signals. To test the stability of the memory states as a function of time, the electronic system was pre-programmed into a conducting (dehydrogenated) or insulating (hydrogenated) state applying $V_t + V_b$ = -2.8 V and +2.8 V, respectively. The retention of the insulating state was measured for over a day with a constant 1N1 = IN2 = 0 V and a reading in-plane drain-source voltage of 0.5 mV was applied for 20 s every 1000 s. During logic-and-memory measurements, the electronic system was pre-programmed into a conducting or insulating state as described above. We then applied the input signals. The optimal parameters were found to be 0.0 V and +1.0 V for both IN1 and IN2 signals, because this yields high $E$, but low $n$, and thus enables strong modulation of the proton channel with minimum disruption of the electronic memory state. We found that in these measurements, the potentials at which graphene becomes hydrogenated are somewhat larger than in our transport maps. We attribute this to the fact that the fast sweeping of the gates may be altering the composition of the electrochemical double layer, probably resulting in lower concentrations of protons in the graphene/electrolyte interface and thus requiring higher potentials to hydrogenate graphene within the fast timescales of this measurement. To implement the logic-and-memory application, the input signals were applied as a function of time in squared waveform patterns. Low and high gate voltages were defined as the logic input 0 and 1, respectively, yielding continuous cycles of different input combinations (00, 01, 11, 10).

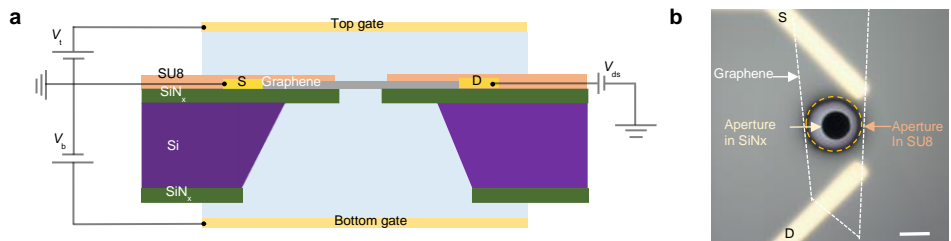

**Extended Data Figure 1| Experimental devices. a**, Schematic of experimental devices used in this work. **b,** Optical image of devices (top view). Dashed white lines mark the area covered by graphene. Dark circle, aperture in silicon-nitride substrate. Dashed orange circle, aperture in the SU-8 washer. All the area shown in the panel (except for the aperture in SU-8 washer) is covered with the washer. S, D labels mark the source and drain electrodes. Scale bar, 10 μm.

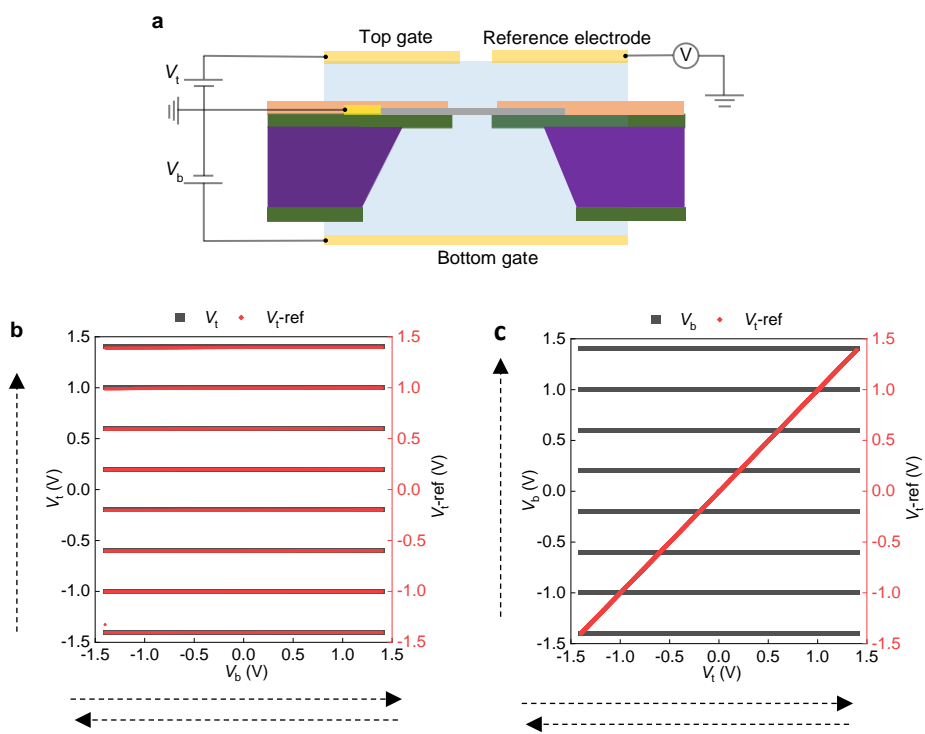

**Extended Data Figure 2| Independence of top and bottom gates. a**, Schematic of experimental devices with reference electrode. **b,** Voltage measured in the reference electrode, $V_t^{ref}$ (red data points), as a function of $V_b$ for a fixed $V_t$ (black data points). The horizontal dashed arrows in the bottom axis indicate that the bottom gate is swept from -1.4 V to and from 1.4 V repeatedly for a fixed value of $V_t$. Vertical dashed line next to the left y-axis indicates that $V_t$ is stepped from -1.4 V to 1.4 V. Left y-axis (black), applied $V_t$. Right y-axis (red), measured $V_t^{ref}$. **c**, Voltage measured in the reference electrode, $V_t^{ref}$ (red data points), as a function of $V_t$ for a fixed $V_b$ (black data points). The horizontal and vertical lines next to x- and y-axis indicate the sweeping and stepping of the gates. Left y-axis (black), applied $V_b$. Right y-axis (red), measured $V_t^{ref}$.

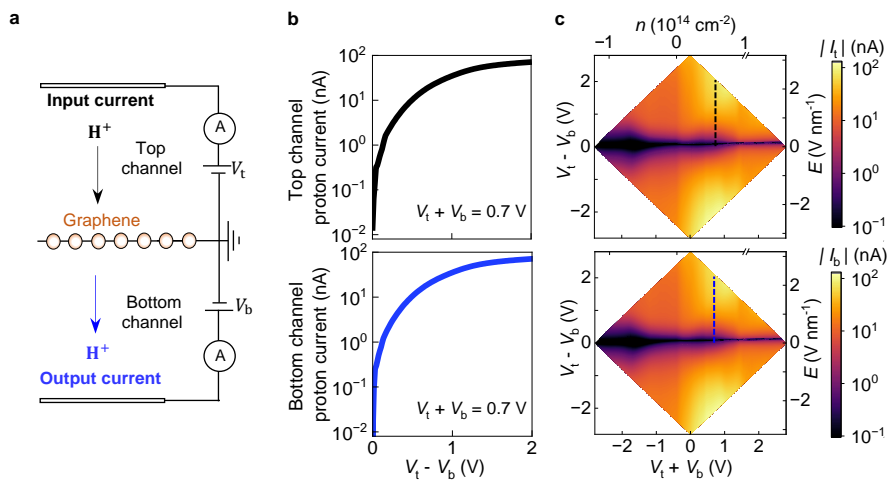

**Extended Data Figure 3| Top and bottom proton channels are symmetry to each other**. **a**, Schematic of devices illustrating the top and bottom channel current. **b**, Top and bottom channel proton transport current taken from the black and blue cross sections in the maps in panel **c** ($V_t + V_b = 0.7$ V), respectively. **c**, Maps of top (top panel) and bottom channel (bottom panel) current as a function of $V_t + V_b$ and $V_t - V_b$.

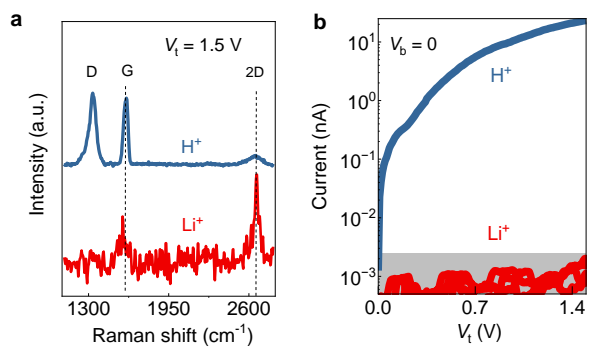

**Extended Data Figure 4| Reference experiments with Li$^+$ conducting electrolyte. a**, Raman spectra for monolayer graphene devices in which free protons are exchanged for Li$^+$ ions do not display a *D* band even at high applied gate voltage, demonstrating that the adsorbed species are indeed protons. The background signal from the electrolytes were subtracted and the spectrum of hydrogenated graphene was divided by a factor of 5 for clarity. **b**, Monolayer graphene is completely impermeable to Li$^+$. Grey area, experimental resolution background.

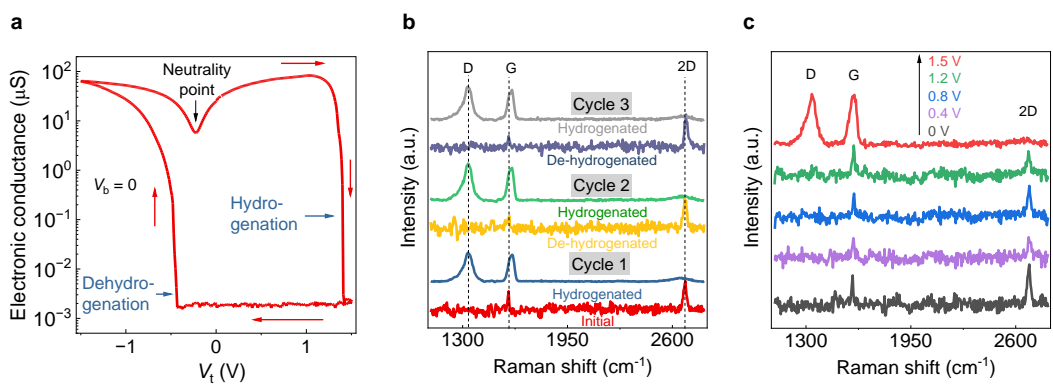

**Extended Data Figure 5| Reversibility of hydrogenation of graphene. a**, In-plane electronic conductance in graphene as a function of gate voltage $V_t$ for $V_b = 0$. The electronic system undergoes a reversible conducting-insulating transition as $V_t$ is swept along the loop marked with red arrows. Blue arrows mark the points at which the system undergoes the reversible insulating (hydrogenation) and conductive (dehydrogenation) transitions. The neutrality point is indicated with a black arrow. Drain-source bias, 0.5 mV. **b**, Raman spectra show that the conducting-insulation transition is accompanied by a sharp *D* band, consistent with hydrogenation of the lattice. The devices can be hydrogenated and dehydrogenated multiple times. Dashed lines mark the position of the *D*, *G* and 2*D* bands. **c**, Raman spectra as a function of $V_t$ show that the *D* band appears suddenly for gate voltages between 1.2 V - 1.5 V. The background signal from the electrolyte was subtracted and the spectra of hydrogenated graphene in panels **b** and **c** were divided by a factor of 5 for clarity.

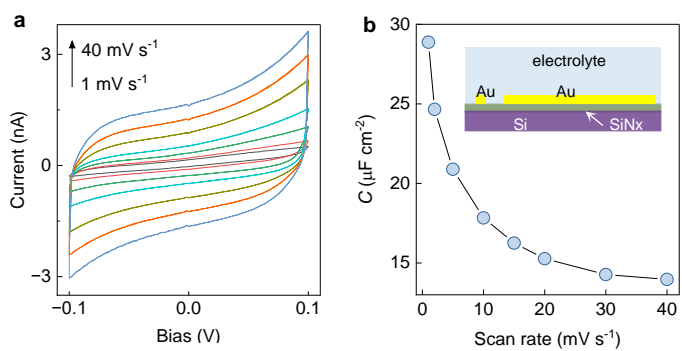

**Extended Data Figure 6| Characterisation of electrolyte capacitance. a**, Cyclic voltammetry characteristics of reference devices in which HTFSI electrolyte is in contact with two mm-sized Au electrodes (inset panel b). The different curves (colour coded) were obtained at sweep rates ranging from 1 mV s$^{-1}$ (black) to 40 mV s$^{-1}$ (dark blue). **b**, Geometrical capacitance per unit area as a function of sweep rate extracted from the CV curves shown in panel **a**. Inset, schematic of reference devices used for this experiment. The dimensions of the active area of the electrodes were 4 mm × 4 mm and 0.3 mm × 1 mm.

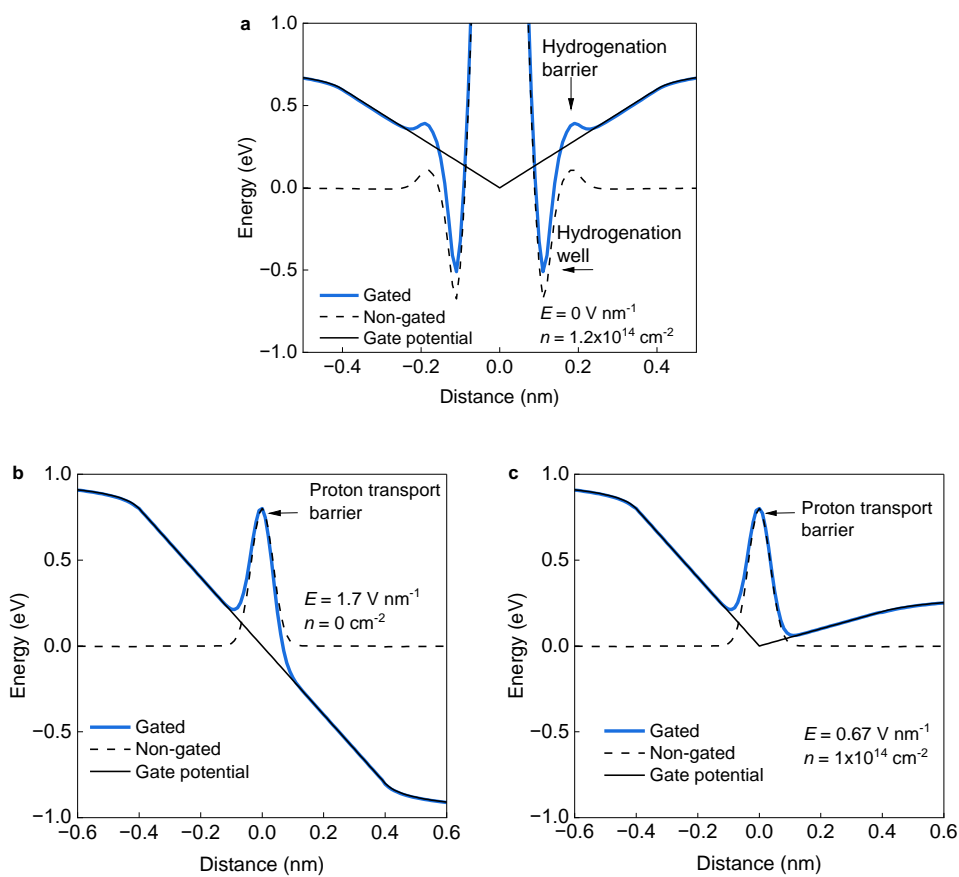

**Extended Data Figure 7| Analytical model of proton transport and hydrogenation in double-gated graphene. a**, High electron doping effectively removes the energy barrier for hydrogenation. Energy profile for graphene hydrogenation in the absence of a gate (dotted curve), the gate potential (solid black curve) and their superposition (blue curve). $V_t = V_b = 0.74$ V. **b**, High electric field lowers the effective energy barrier for proton transport. Energy profile for proton transport through graphene in the absence of a gate (dotted curve), the gate potential (solid black curve) and their superposition (blue curve). $V_t = - V_b = 0.96$ V. **c**, High electron doping lowers the effective energy barrier for proton transport. Energy profile for proton transport through graphene in the absence of a gate (dotted curve), the gate potential (solid black curve) and their superposition (blue curve). $V_t = 0.96$ V, $V_b = 0.29$ V.

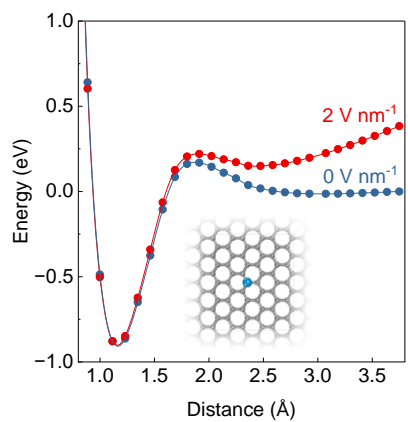

**Extended Data Figure 8| Calculated potential energy profile for graphene hydrogenation.** Potential energy vs distance for a proton over a carbon atom in graphene. An electric field perpendicular to the graphene sheet distorts the potential energy profile (red data points), with respect to the case where no field is applied (blue data points). Red and blue curves, spline interpolation to data. Inset, schematic of a proton (blue ball) on top of the graphene lattice (grey balls).

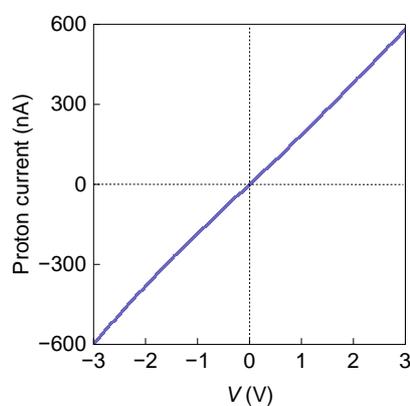

**Extended Data Figure 9| Characterisation of reference devices without graphene. a**, Examples of *I-V* characteristics of 'open hole' device. The device consists of a 10 μm diameter hole etched in a silicon nitride substrate; HTFSI electrolyte on both sides; and two $PdH_x$ electrodes. Dashed lines, guide to the eye.

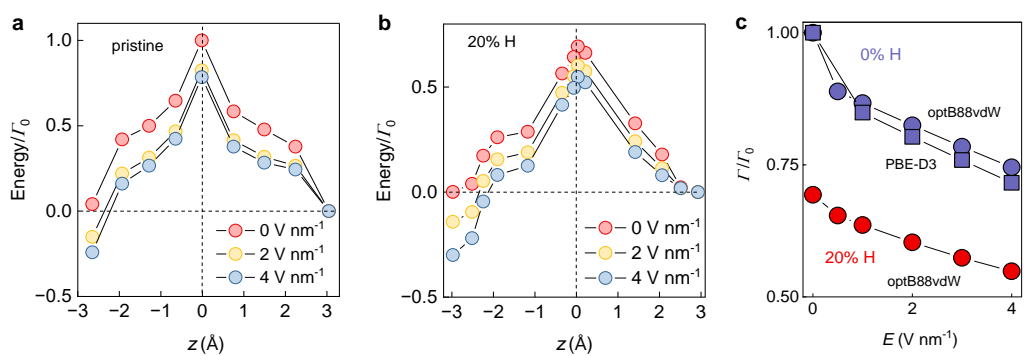

**Extended Data Figure 10| Calculated energy barriers for proton transport through graphene. a**, Energy barriers for proton transport through graphene under different electric fields for no hydrogenation and **b**, for 20% H atom coverage, normalised vs $\Gamma_0$, the barrier height for the case of no H atoms absorption and zero electric field. **c**, DFT calculations show that $E$ and H atom adsorption lower the energy barrier for proton transport, $\Gamma$. Red (blue) symbols, $\Gamma/\Gamma_0$ for the case of no H atom adsorption (20% H coverage). Square and circle symbols, data obtained using the optB88-vdW and PBE-D3 functionals, respectively.